\documentclass[aps,prl,preprint,superscriptaddress,nofootinbib,showkeys,floatfix]{revtex4}
\usepackage{amsmath,amssymb,graphicx}
\begin{document}
\preprint{\parbox[b]{1in}{ \hbox{\tt PNUTP-14-A02} }}
%\preprint{}
\newcommand{\upperRomannumeral}[1]{\uppercase\expandafter{\romannumeral#1}}
\newcommand{\lowerromannumeral}[1]{\romannumeral#1\relax}
\title{Holographic Estimate of Electromagnetic Mass}


%\author{D.~K. Hong}
\author{Deog Ki  Hong}
\email[]{dkhong@pusan.ac.kr}
\affiliation{Department of
Physics,   Pusan National University,
             Busan 609-735, Korea}
% \affiliation{Asia Pacific Center for Theoretical Physics,  POSTECH, Pohang 709-784, Korea}            


\date{\today}

\begin{abstract}
Using the gauge/gravity duality, we calculate the electromagnetic contributions to hadron masses, where mass generates dynamically by strong QCD interactions. Based on the Sakai-Sugimoto model of holographic QCD we find that the electromagnetic mass of proton is $0.48~{\rm MeV}$ larger than that of neutron, which is in agreement with recent lattice results.  Similarly for pions we obtain $m_{\pi^{\pm}}-m_{\pi^0}=1.8~{\rm MeV}$, roughly half of the experimental value.
The electromagnetic mass of pions is found to be independent of $N_c$ and  't Hooft coupling and its scale is set only by the Kaluza-Klein scale of the model, $M_{\rm KK}=949~{\rm MeV}$. 
 
\end{abstract}


\maketitle

\section{Introduction}
\label{intro}
Quantum chromodynamics (QCD) is now firmly accepted as the theory of strong interactions. It is well tested not only at high energy, where QCD couples weakly,  but also at low energy, where QCD becomes strongly interacting. In practice, however,  QCD is notoriously hard to solve directly to describe  the properties of hadrons such as mass spectrum, form factors, or other matrix elements, which are intrinsically nonperturbative.  Understanding the low-energy physics of hadrons in terms of QCD still remains to be a great challenge. In recent years, however, the lattice QCD has progressed enormously to simulate with parameters such as the light quark masses very close to physical values in describing the properties of hadrons, whose analytic understanding is therefore pressing. 

Another development to understand low-energy QCD has been made recently in string theory. Recent study of string theory has shown that a strongly coupled conformal field theory (CFT) is dual to a weakly interacting gravity in anti-de Sitter (AdS) space in the limit of  large degrees of freedom, which is known as AdS/CFT correspondence. This finding leads to a new possibility of solving QCD in a certain limit, if one finds the holographic dual of QCD.
  
Several models for holographic QCD were proposed and were shown to be quite successful in describing the properties of hadrons~\cite{Sakai:2004cn,Sakai:2005yt,Erlich:2005qh,Da Rold:2005zs,Hong:2006ta}.  Holographic QCD deals directly with hadrons as basic degrees of freedom and the dynamics of hadrons is determined by the gauge/gravity duality prescription, found in string theory when both the number of colors, $N_c$, and the 't Hooft coupling, $\lambda=g^2N_c$, are very large. 

In this paper we calculate the electromagnetic contributions to the hadron masses in the Sakai-Sugimoto model of holographic QCD~\cite{Sakai:2004cn}, which is constructed from a type \upperRomannumeral{2}A string theory with D4-D8 branes to  provide a successful effective theory of hadrons, consistent with QCD~\cite{Sakai:2005yt}. 
Calculating accurately  the electromagnetic mass is quite important, since it is vital to determine the precise value of the light quark masses from the hadron spectrum.
Fortunately, the lattice calculation on the electromagnetic contributions to hadron mass has been progressed  a lot to accurately determine the electromagnetic mass of hadrons~\cite{Blum:2007cy,Basak:2008na,Blum:2010ym,deDivitiis:2013xla,Borsanyi:2013lga,Borsanyi:2014jba}. In this paper we provide a holographic estimate of the electromagnetic mass of hadrons, which will be  complementary to the lattice results.

\section{Meson Masses}
One of the successes of QCD is that it naturally explains why low-lying  mesons like pions and kaons are much lighter than baryons. 
QCD with three (light) flavors  has a chiral symmetry, ${\rm SU}(3)_L\times {\rm SU}(3)_R$ in the massless limit, which  is predicted to be spontaneously broken to ${\rm SU}(3)_V$ due to the strong dynamics of QCD. Pions and kaons are then identified as the Nambu-Goldstone bosons associated with the spontaneously broken chiral symmetry. By Goldstone theorem, the Nambu-Goldstone bosons are massless when the chiral symmetry is exact. However, the chiral symmetry is only approximate in QCD due to the current quark mass and also electroweak interactions, which then makes pions and kaons massive, but much lighter than baryons.

When a continuous global  symmetry, $G_f$, is spontaneously broken to its subgroup $H$ and also intrinsically broken to $S_w\subset G_f$, the orientation of the true vacuum should be selected or aligned to minimize the vacuum energy, induced by the intrinsic breaking. Provided that the intrinsic breaking is perturbative, the potential for the vacuum orientation $g\in G_f$ is given to the lowest order as
\begin{equation}
V(g)=\left<0,g\right|{\cal H}^{\prime}\left|0,g\right>,
\end{equation}
where $\left|0,g\right>=U(g)\left|0\right>$ is a $G_f$-rotated vacuum of QCD and ${\cal H}^{\prime}$ is the Hamiltonian that breaks $G_f$ intrinsically. Instead of rotating the vacuum, one might rotate the external perturbation, 
\begin{equation}
V(g)=\left<0\right|U(g)^{-1}{\cal H}^{\prime}U(g)\left|0\right>,
\end{equation} 
then the minimization of the vacuum energy becomes tantamount to finding the direction of the symmetry-breaking Hamiltonian in the basis of the unperturbed vacuum.
Finding the vacuum alignment due to the intrinsic breaking was studied by Dashen~\cite{Dashen:1970et} for the chiral symmetry breaking in strong interactions and subsequently by Weinberg~\cite{Weinberg:1975gm} in a generalized context of dynamical electroweak symmetry breaking. Later it  has been applied to calculate masses of Nambu-Goldstone bosons in technicolor~\cite{Peskin:1980gc,Preskill:1980mz}. 

There are two sources for the intrinsic breaking of chiral symmetry in QCD. One is current quark mass and the other is the electroweak interaction. At the leading order in perturbation the vacuum alignment by both current quark mass and electroweak interactions can be treated independently.   In this paper we study the vacuum alignment due to the electromagnetic interaction and calculate the electromagnetic corrections to the pion mass in holographic QCD. The vacuum alignment in holographic QCD due to the current quark mass has been studied previously~\cite{Aharony:2008an,Hashimoto:2008sr,Hong:2007tf}.

\begin{figure}[tbh]
	\centering
	\includegraphics[width=0.5\textwidth]{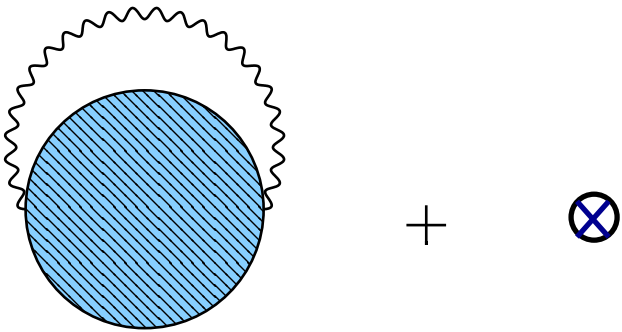}%
		\caption{\label{vac_e}The electromagnetic corrections to the vacuum energy at the leading order. The blob denotes the vacuum polarization and the wiggly line denotes the photon propagator. A counter term to the vacuum energy is denoted as $\otimes$, which is independent of the pion fields.}
\end{figure}

The vacuum energy density due to the electromagnetic interaction is given in the leading order as (See Fig.~\ref{vac_e})
\begin{equation}
{\cal E}_{\rm vac}=-\frac{e^2}{2}\int{\rm d}^4x\,D^{\mu\nu}(x)\left<0\right|U^{-1}J_{\mu}^{Q_{\rm em}}(x)J_{\nu}^{Q_{\rm em}}(0)\,U\left|0\right>\,,
\end{equation}
where  $U=e^{2i\pi/f_{\pi}}$ describes the pion fields, $J_{\mu}^{Q_{\rm em}}$ the electromagnetic currents of quarks, and $D^{\mu\nu}(x)$ denotes the photon propagator. The electric charge operator, $Q_{\rm em}$, in the basis of the strong-interaction vacuum, is defined as a sum of isospin and hypercharge, $Q_{\rm em}=I_3+\frac12Y$ up to a $G_f$-rotation. 

For simplicity, we consider two-flavor QCD, where $G_f={\rm SU}(2)_L\times {\rm SU}(2)_R\times {\rm U(1)}_V$ and $Y$ becomes the baryon number. 
The electromagnetic (EM) mass of pions is now given as~\cite{Peskin:1980gc,Preskill:1980mz} 
\begin{equation}
m_{\pi^{\pm}}^2-m_{\pi^0}^2=\left.\frac{\partial^2}{\partial \pi_+\partial \pi_-}{\cal E}_{\rm vac}[U]\right|_{U=1}\equiv e^2M^2,
\end{equation}
where 
\begin{equation}
M^2=\frac{1}{f_{\pi}^2}\int{\rm d}^4x\,D^{\mu\nu}(x)\left<0\right| T \left[V_{\mu}^3(x)V^3_{\nu}(0)-A_{\mu}^3(x)A_{\nu}^3(0)\right]\left|0\right>,
\end{equation}
with the vector and axial vector flavor currents, $V_{\mu}^a$ and $A_{\mu}^a$ ($a=1,2,3$) respectively. 

The vacuum correlator  of (vector) flavor currents is given as 
\begin{equation}
\Pi_{V\mu\nu}^{ab}(q)=\int {\rm d}^4x\,e^{iq\cdot x}\left<0\right|V_{\mu}^a(x)V_{\nu}^b(0)\left|0\right>
=\delta^{ab}\left(q_{\mu}q_{\nu}-q^2g_{\mu\nu}\right)\Pi_V(q)\,,
\end{equation}
and similarly the vacuum correlator of axial vector currents is given as  
\begin{equation}
\Pi_{A\mu\nu}^{ab}(q)=\int {\rm d}^4x\,e^{iq\cdot x}\left<0\right|A_{\mu}^a(x)A_{\nu}^b(0)\left|0\right>
=\delta^{ab}\left(q_{\mu}q_{\nu}-q^2g_{\mu\nu}\right)\Pi_A(q)+\frac{q^{\mu}q^{\nu}}{q^2}f_{\pi}^2\delta^{ab}\,.
\end{equation}
The vacuum correlators are nonperturbative and has precluded any analytic calculations, though several attempts were made to estimate the vacuum correlator by using QCD sum rules or dispersion relations among others. 
However, 
in the large $N_c$ and large $\lambda$ limit, where  QCD is described by its gravity dual, called holographic QCD, 
the correlators of flavor currents are easily calculated by gauge/gravity duality. The holographic QCD is  described in general by a five-dimensional Chern-Simons-Yang-Mills theory for the flavor symmetry of QCD, the boundary gauge theory, 
\begin{align}
S_A=\kappa\int{\rm d}^5x\,{\rm Tr}\left[-\frac{1}{2g^2(z)}F_{\mu\nu}^2+f^2(z)F_{z\mu}^2\right]+S_{\rm CS}\,,
\label{action}
\end{align}
where $S_{\rm CS}=N_c/24\pi^2\int\omega_5(A)$ is the 5D Chern-Simons action, that reproduces the QCD flavor anomaly, and the fifth coordinate $z$ corresponds to the holographic direction with a warp factor, parameterized by $g(z)$ and $f(z)$. 
In the case of Sakai-Sugimoto model, that we will focus on in the paper, $g^3=f=\sqrt{1+z^2}$\,, taking the Kaluza-Klein scale $M_{\rm KK}=949~{\rm MeV}$ to be the unit of the scale, and $\kappa=N_c\,\lambda/(216\pi^3)$.
Another popular approach to holographic QCD is a so-called bottom-up approach, first introduced in~\cite{Erlich:2005qh,Da Rold:2005zs}. The mass difference of $\pi^{\pm}$ and $\pi^0$ in the hard-wall model was calculated in~\cite{Da Rold:2005zs}.
By the AdS/CFT correspondence the 5D action becomes the generating functional for one-particle irreducible Green's function of 4D operators in the limit of  large number of color ($N_c\gg1$) and large 't Hooft coupling ($\lambda\gg1$), if evaluated on-shell for the bulk fields dual to the 4D operators, whose ultra-violet values are identified as the sources of the operators.

\begin{figure}[tbh]
	\includegraphics[width=0.28\textwidth]{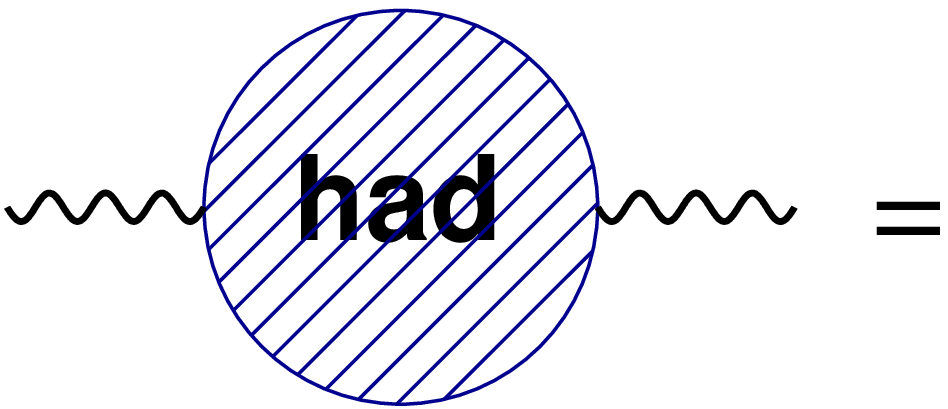}%
	\includegraphics[width=0.5\textwidth]{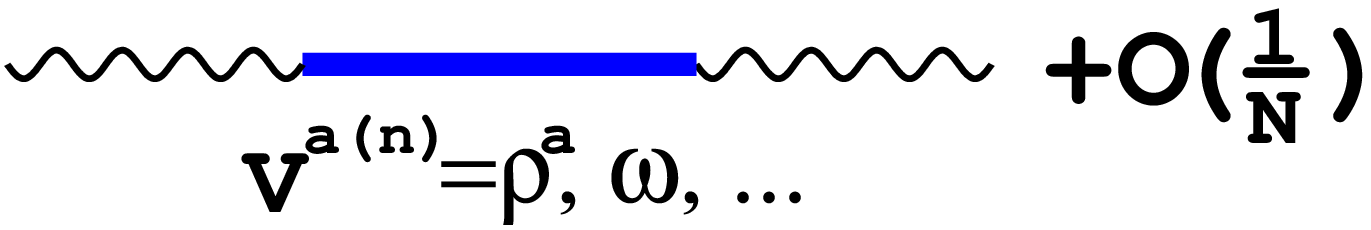}
	\caption{\label{pol}The vacuum polarization in holographic QCD.}
\end{figure}

By the AdS/CFT prescription the (axial) vector current-current correlators in the momentum space are given by the bulk action evaluated on-shell as
\begin{equation}
\Pi_{V(A)}(q^2)=-\left.\frac{2\kappa}{q^2}\left(1+z^2\right)\partial_z\,A(z,q^2)\right|_{z_m},
\label{vac_pol}
\end{equation} 
where $z_m$ is the UV boundary of the holographic coordinate that goes to infinity in the Sakai-Sugimoto model and $A(z,q^2)$ is the non-normalizable bulk gauge field, up to the polarization vector, that satisfies the equation of motion,
\begin{equation}
(1+z^2)^{4/3}\,\partial_z^2A(z,q^2)+2z(1+z^2)^{1/3}\,\partial_zA(z,q^2)+q^2A(z,q^2)=0\,
\end{equation}
with (anti-) symmetric UV boundary conditions for (axial) vector currents. 
If we expand the non-normalizable modes, corresponding to the source, in the basis of normalizable modes of bulk solutions as a sum of infinite tower of (axial) vector mesons (see Fig.~\ref{pol}), using the decomposition formula~\cite{Hong:2004sa}, we get for the vector and axial vector correlators, respectively, following the notations in~\cite{Sakai:2005yt},
\begin{equation} 
   \Pi_V(q^2) = 
\sum_{n=1}^\infty \frac{g_{v^n}^2}{(q^2 -  m_{v^n}^2) m_{v^n}^2}\,,\quad 
 \Pi_A(q^2) = -\frac{f_{\pi}^2}{q^2}+
\sum_{n=1}^\infty \frac{g_{a^n}^2}{(q^2 -  m_{a^n}^2) m_{a^n}^2}\,,
     \label{piv:res}
\end{equation}
where $g_{v^n}\,(g_{a^n})$ and $m_{v^n}\,(m_{a^n})$ are the decay constants and mass of $n$-th (axial) vector mesons, respectively. 

After Wick-rotation, the electromagnetic mass  becomes, generalizing the formula by  Das {\it et al.}~\cite{Das:1967it},
\begin{eqnarray}
e^2M^2&=&\frac{3e^2}{f_{\pi}^2}\int\frac{{\rm d}^4Q}{(2\pi)^4}\left[\Pi_V(-Q^2)-\Pi_A(-Q^2)\right]\label{integral}\\
&=&\frac{3e^2}{8\pi^2f_{\pi}^2}\sum_n\left[
g_{v^n}^2\ln\left(\frac{\Lambda}{m_{v^n}}\right)-g_{a^n}^2\ln\left(\frac{\Lambda}{m_{a^n}}\right)\right]\,,\label{pion}
\end{eqnarray}
where  $\Lambda$ is the UV cutoff.  The Weinberg first sum rule on spectral functions is used in the second line, since  (axial) vector currents are conserved in hQCD\footnote{We also checked numerically that  $\lim_{Q^2\to0}Q^2\Pi_A(-Q^2)=f_{\pi}^2$ and $\lim_{Q^2\to0}Q^2\Pi_V(-Q^2)=0$\, to find the Weinberg first sum rule holds in the Sakai-Sugimoto model.}.
Being a radiative correction to the pion mass term, the EM corrections could be quadratically divergent. But, the quadratic divergence is absent due to the Weinberg first sum rule. 
We further note that, 
since  the chiral symmetry is spontaneously broken to vector symmetry in the holographic QCD, the Weinberg second sum rule on the spectral functions, $\sum_n(g^2_{v^n}-g^2_{a^n})=0$ should hold in hQCD as well~\cite{Weinberg:1996kr}. 
Therefore, though each term in the sum of Eq.~(\ref{pion}) is logarithmically divergent,  the logarithmic divergence is absent in the sum and the EM mass of pions is finite and thus independent of the UV cutoff.

%the UV divergent piece in the EM corrections is independent of the pion decay constant, $f_{\pi}$ and present even when pions are massless. Since, however, the pions become massless and decouple in the limit $f_{\pi}\to\infty$, we demand that the counter term $\delta m_{\pi}^2$ cancels the UV divergence exactly\footnote{ }. 

\begin{figure}[tbh]
	\includegraphics[width=0.43\textwidth]{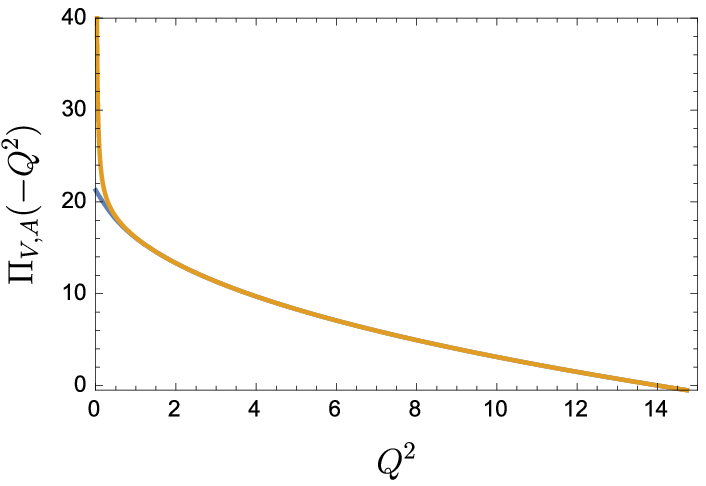}\hskip 0.2in
	\includegraphics[width=0.45\textwidth]{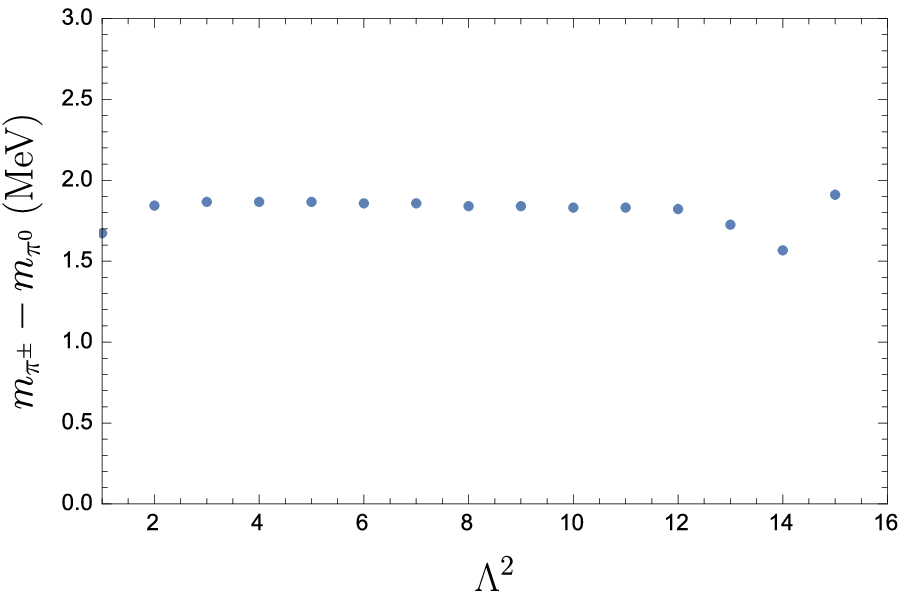}
	\\ \hskip 0.3in (a) \hskip 2.5in (b)
	\caption{\label{cor}(a) The vector and axial vector correlation functions. The orange line denotes $\Pi_A(-Q^2)$ and the blue line $\Pi_V(-Q^2)$. (b) The pion EM mass. ($Q^2$ and $\Lambda^2$ are in the unit of $M_{\rm KK}^2$.) }
\end{figure}

We find numerically the vector and axial vector correlation functions converge to each other rather quickly in the Sakai-Sugimoto model (See Fig.~\ref{cor}a). Though the EM mass of pions is finite, we introduce a UV cutoff $\Lambda$ for the numerical estimate of the integral, (\ref{integral}). However, as shown in Fig.~\ref{cor}b, the electromagnetic mass~\footnote{Since $f_{\pi}^2=4\kappa/\pi\,M_{\rm KK}^2$, the EM mass of pions is independent of $N_c$ and $\lambda$ and its scale is set only by $M_{\rm KK}=949~{\rm MeV}$.} of pion converges rather quickly to $1.8~{\rm MeV}$ for $\Lambda^2>2M_{\rm KK}^2$ and insensitive to the UV cutoff. We estimate, therefore, the pion EM mass in the Sakai-Sugimoto model to be $1.8~{\rm MeV}$, which is less than half of the experimental value, $4.5~{\rm MeV}$. Compared to the result obtained by Das {\it et al.}~\cite{Das:1967it}, $5~{\rm MeV}$, and also to the hard wall calculation~\cite{Da Rold:2005zs}, $3.6~{\rm MeV}$, our estimate of the pion EM mass in the Sakai-Sugimoto model turns out to be rather small.

\section{Baryons in holographic QCD}

Baryons in the large $N_c$ QCD should be realized as solitons~\cite{Witten:1979kh}. In holographic QCD, being a 5D gauge theory, there is a topologically conservered current 
\begin{equation}
J^{M}=\frac{1}{32\pi^2}\epsilon^{MNLPQ}\,{\rm tr}\left(F_{NL}F_{PQ}\right)\,,
\end{equation}
where the 5d coordinates $M,N,L,P,Q=0,1,2,3,z$\,.
As the 5d current couples to the $U(1)$ bulk gauge field, corresponding to the quark number, through the 5d Chern-Simons term, one can define the 4d baryon current as
\begin{equation}
B^{\mu}=\frac{1}{8\pi^2}\int \!{\rm d}z\,\epsilon^{\mu\nu\rho\sigma}{\rm tr}\left(F_{\nu\rho}F_{\sigma\,z}\right),
\end{equation}
which becomes the Skyrme current upon the Kaluza-Klein reduction of the bulk gauge fields~\cite{Hong:2008nh}.  
Baryons are hence realized as topological solitons in holographic QCD~\cite{Hong:2007kx,Hong:2007ay,Hata:2007mb}, whose topological charge is nothing but the instanton number, identified as baryon numbers;
\begin{equation}
\int\!{\rm d}^3x\,{\rm d}z\,J^0(x,z)=\frac{1}{32\pi^2}\int\!{\rm d}^3x\,{\rm d}z\,F_{IJ}^a{\tilde F}^{aIJ}=\int\! {\rm d}^3x B^0(x)=N_B\,,
\end{equation}
where $F_{IJ}^a$ are the field strength tensors of the bulk flavor gauge fields and ${\tilde F_{IJ}}^a$ are their dual with $I,J=1,2,3,z$\,. 

In holographic QCD the size of instanton is not a zero mode but has a potential, due to the warped geometry and to the U(1) Coulomb repulsion, which has a minimum away from the origin. 
Since the Coulomb energy is coming from the CS term, which is subleading in $1/{\lambda}$-expansion, compared to the DBI energy, the size of soliton is expected to be of the order of $1/\sqrt{\lambda}$, quite small in the large $\lambda$ limit.

The energy of the soliton from the DBI part of the 5d action (\ref{action}) is given as 
\begin{equation}
{\cal E}_0=\kappa\int{\rm d}^3x{\rm d}z\left[\frac{1}{4}\left(1+z^2\right)^{-1/3}{\left(F_{ij}^a\right)}^2+\frac12\left(1+z^2\right)\left(F_{iz}^a\right)^2\right]\,,
\end{equation}
where the SU(2) index $a=1,2,3$ and the spatial index $i,j=1,2,3$\,.
Since the instanton soliton satisfies the duality condition,
\begin{equation}
F_{iz}^a=\frac12\sqrt{-g_4}\,\epsilon_{izlk}\,F^{alk},
\end{equation} 
where $\epsilon$ is the fully antisymmetric tensor in the flat space and $g_4$ is the determinant of the induced 4d metric of the $z={\rm constant}$ hyper surface,
we find that 
\begin{equation}
F_{iz}^a=\frac12\left(1+z^2\right)^{-2/3}\epsilon_{ijk}F_{jk}^a\,.
\end{equation}
The DBI energy can be then written as
\begin{equation}
{\cal E}_0={\kappa}\int{\rm d}^3x\,{\rm d}z\left(1+z^2\right)^{1/3}{\vec E^a}\cdot{\vec B^a}\,.
\end{equation}
where $E_i^{a}=F_{iz}^a$ and $B_i^a=\frac12\epsilon_{ijk}F_{jk}^a$\,.
For the small size soliton, located at the origin $z=0$, we may use the flat-space instanton soluton with an instanton density 
for $N_B=1$
\begin{equation}
\frac{1}{8\pi^2}\,{\vec E^a}\cdot{\vec B^a}=\frac{6}{\pi^2}\cdot\frac{\rho^4}{\left(\rho^2+{\vec x}^2+z^2\right)^4}\,,
\end{equation}
where $\rho$ is the size of the instanton. For $\rho\ll1\,(\equiv M_{\rm KK}^{-1})$ the DBI energy may be expanded in powers of $\rho$ to get 
\begin{equation}
{\cal E}_0=m_B^{(0)}\left(1+\frac16\,\rho^2+{\cdots}\right)\,
\end{equation}
where $m_B^{(0)}=8\pi^2\kappa=\lambda N_c\,M_{\rm KK}/(27\pi)$.
Without the Coulomb repulsion the potential for the size $\rho$ will have minimum at the origin so that the instanton soliton is unstable against shrinking to a zero size. 
However, since the instanton sources through the 5D Chern-Simons term the $U(1)$ gauge field in the bulk, the Coulomb repulsion balances the gravitational attraction to stabilize the instanton soliton. 
The action for the bulk $U(1)$ can be written as 
\begin{equation}
S_{U(1)}=\kappa\int\! {\rm d}^4x{\rm d}z\left[-\frac14\left(1+z^2\right)^{-1/3}F_{\mu\nu}^2+\frac12\left(1+z^2\right)F_{\mu z}^2\right]+\frac12N_c
\int\! {\rm d}^4x{\rm d}z\,A_0J^0\,,
\end{equation}
where the electrostatic potential energy is coming from the 5d Chern-Simons term of holographic QCD action (\ref{action}). 
For a given, static $U(1)$ charge distribution $J^0(x_i,z)$,
its Coulomb energy is given as 
\begin{eqnarray}
{\cal E}_{\rm C}&=&\kappa\int\! {\rm d}^3x{\rm d}z\left[-\frac12 \left(1+z^2\right)^{-1/2}E_i^2-\frac12\left(1+z^2\right)E_z^2\right]+\frac12N_c\int\! {\rm d}^4x{\rm d}z\,A_0J^0\\
&=&\frac14N_c\int\! {\rm d}^3x{\rm d}z\,A_0J^0\,,
\end{eqnarray}
where we used the Gauss's law in the second line. For the small size soliton located at the origin
\begin{equation}
J^0(x,z)\simeq\frac{6}{\pi^2}\cdot\frac{\rho^4}{\left(\rho^2+{\vec x}^2+z^2\right)^4}\,,
\end{equation} 
the electric field can be obtained by the Gauss's law
\begin{equation}
\int \vec E\cdot {\rm d}\vec \Sigma=\frac{N_c}{2\kappa}\cdot\frac{6}{\pi^2}\int_0^r\frac{\rho^4\, 2\pi^2
r^3{\rm d}r}{\left(\rho^2+r^2\right)^4}\,,
\end{equation}
from which we get 
\begin{equation}
E_r=\frac{N_c}{4\pi^2\kappa\, r^3}\left[1-\frac{\rho^4\left(\rho^2+3r^2\right)}{\left(\rho^2+r^2\right)^3}\right]\,.
\end{equation}
Therefore, the Coulomb energy becomes in the expansion of powers of $\rho$
\begin{equation}
{\cal E}_C\simeq\frac{\kappa}{2}\int\!{\rm d}^3x\,{\rm d}z\,E_r^2=\frac{N_c^2}{40{\pi}^2\kappa}\cdot\left(\frac{1}{\rho^2}+{\cal O}(1)\right).
\end{equation}
The total energy of the instanton soliton becomes for $\rho\ll1$
\begin{equation}
{\cal E}=8\pi^2\kappa+\frac{4\pi^2\kappa}{3}{\rho^2}+\frac{N_c^2}{40{\pi}^2\kappa}\cdot\frac{1}{\rho^2}+\cdots\,.
\label{energy}
\end{equation}
We see that the size of the instanton soliton, that minimizes the total energy (\ref{energy}), becomes in the unit of $M_{KK}^{-1}$
\begin{equation}
\rho_B\simeq \frac{1}{\pi}\left(\frac{3}{40}\right)^{1/4}\sqrt{\frac{N_c}{2\kappa}}=\frac{9.6}{\sqrt{\lambda}}\,,
\end{equation}
which is quite small in the large $\lambda$ limit, as the Coulomb potential energy is subdominant in the $1/\lambda$ expansion, compared to the DBI energy~\cite{Hong:2007kx,Hong:2007ay,Hata:2007mb} . Being a small-sized soliton, the supergravity approximation of Sakai-Sugimoto model might be invalid to describe holographic baryons. However, it has been shown that the stringy effects to the instanton solitons are numerically suppressed~\cite{Hashimoto:2010je}. 

Since the size of instanton solitons  turns out to be much smaller than the characteristic length of holographic QCD,  the holographic baryons can be described by point-like bulk spinor fields at low energy, assuming $N_c$ is odd. Similarly to the bottom-up model of holographic baryons~\cite{Hong:2006ta}, a 5D effective action for the bulk baryon field has been derived in the 5D momentum expansion to be consistent with the bulk gauge symmetry~\cite{Hong:2007kx}. Especially the coefficient of the Pauli term has been calculated for the baryons sitting at the origin, $z=0$, at the leading order, which is essential to correctly reproduce the axial couplings of nucleons among others.  

The (low-energy) effective action for the holographic baryons is given in powers of momentum by the AdS/CFT correspondence as~\cite{Hong:2006ta,Hong:2007kx},  
\begin{equation}
S_B=\int{\rm d}^5x\sqrt{g}\left[\bar B\, e_A^M\Gamma^A\nabla_M B-m_b(z)\bar BB+\mu_5(z)\bar B\,\Gamma^{MN}F_{MN}B+{\rm h.o.}\right],
\label{baryon_eff}
\end{equation}  
where $B$ is the 5d bulk spinor with a position-dependent mass $m_b(z)$, $e_M^A$ is the vielbein, satisfying $g_{MN}=e_M^Ae_N^B\eta_{AB}$, and the Dirac matrices satisfy $\left\{\Gamma^A,\Gamma^B\right\}=2\eta^{AB}$\,. The covariant derivative 
\begin{equation}
\nabla_M=\partial_M+\frac{i}{4}\omega_M^{AB}\Gamma_{AB}-iA_M^at^a\,.
\end{equation}
where the Lorentz generator $\Gamma^{AB}=\frac{1}{2i}\left[\Gamma^A,\Gamma^B\right]$\,. 
The coupling of  Pauli term is determined by the fact that it should produce the correct long distance tail of instanton solitons that the spinor sources. For the spinor located at the origin, $\mu_5(0)=4\pi^2\rho_b^2/(3\kappa)$~\cite{Hong:2007kx}. The effective action gives couplings of baryons to mesons, upon the Kaluza-Klein reduction of the bulk fields, which turn out to be in good agreement with data. 

\section{Electromagnetic masses of Baryons}
\begin{figure}[tbh]
	\centering
	\includegraphics[width=0.6\textwidth]{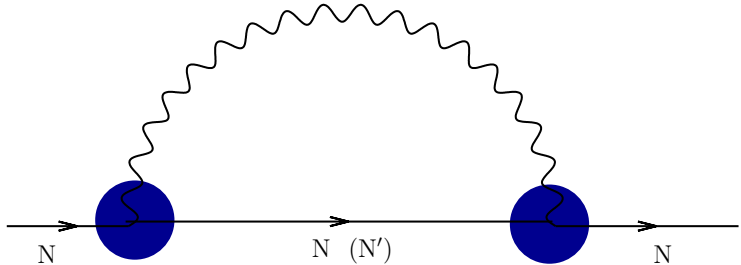}%
		\caption{\label{nucleon}An electromagnetic correction to the nucleon mass at the leading order in $\alpha_{\rm em}$. The blob denotes the EM form-factor of the nucleon. ${\rm N}$ denotes low-lying nucleons and ${\rm N}^{\prime}$ denotes  excited nucleons.}
\end{figure}

\begin{figure}[tbh]
	\centering
	\includegraphics[width=0.8\textwidth]{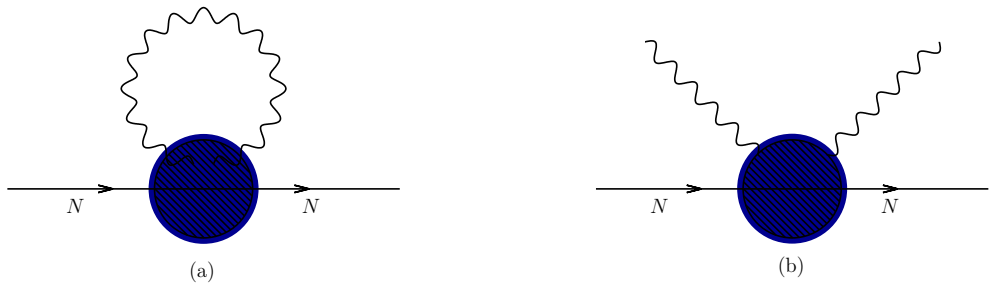}%
		\caption{\label{self}The wiggly line denotes a photon and the solid line the nucleon. (a) An electromagnetic correction to the nucleon self energy. 
		(b) A non-minimal electromagnetic coupling of nucleons.}
\end{figure}

At the leading order in the $\alpha_{\rm em}$ expansion the electromagnetic (EM) contributions to baryon masses come from  two diagrams, shown in Fig.~\ref{nucleon} and Fig.~\ref{self}\,(a).  
The EM correction like the diagram, Fig.~\ref{self}\,(a),  is in general present because baryons have the non-minimal EM couplings,~Fig.~\ref{self}\,(b), as well as the minimal coupling with form factors.

The EM form-factors  of nucleons are in general non-perturbative and difficult to calculate. But, in holographic QCD they are rather easily calculated from the wave-function overlap integration by the AdS/CFT prescription. As was shown in~\cite{Hong:2007kx}, the Pauli term in the non-minimal coupling of bulk spinors in~(\ref{baryon_eff}) does not contain a ${\rm U(1)}$ part, as the spinor sources the instanton solitons, which have only a ${\rm SU(2)}$ long-distance tail.  An immediate consequence of this is that the two-spinor two-photon diagram, Fig~\ref{self}\,(b), vanishes. The EM corrections to the baryon self energy, coming from the diagram in Fig.~\ref{self}\,(a),  therefore vanish as well, not just at the leading order  but at all orders in the $\alpha_{\rm em}$ expansions. In holographic QCD the EM corrections to the baryon mass therefore come only from the diagram, shown in Fig.~\ref{nucleon}. 

\begin{figure}[tbh]
	\centering
	\includegraphics[width=0.25\textwidth]{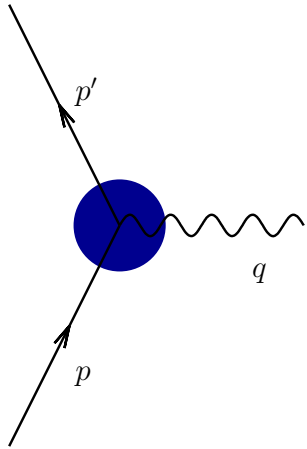}%
		\caption{\label{ff}The blob denotes the electromagnetic (transition) form factors of nucleons.}
\end{figure}

To the leading order we can replace the blob in Fig.~\ref{nucleon} by the Dirac EM form factors of nucleons, since the nucleons in the loop are almost on-shell as the nucleon mass is close to the ultraviolet (UV) cutoff, $M_{\rm KK}\,(=\!949\,{\rm MeV})$ of the holographic effective theory~\footnote{Note that since the form factor decays rather quickly as $1/Q^2$ for the large momentum transfer, the loop diagram, shown in Fig.~\ref{nucleon}, is finite. Furthermore, as the transition form factor, which does not have the Dirac form factor $F_1$, is 
suppressed by the nucleon mass, the contributions of excited nucleons in the loop is negligible.}.
The electromagnetic form factors of nucleons (Fig.~\ref{ff}) are the matrix elements of an external electromagnetic current, which correspond to a non-normalizable ${\rm U}(1)$ gauge field in the bulk with a charge $Q_{\rm em}=I_3+\frac12B$  by the AdS/CFT dictionary:
\begin{equation}
\left<p^{\prime}\right|J_{\rm em}^{\mu}(x)\left|p\right>=e^{iqx}\,\bar
u(p^{\prime})\,{\cal O}^{\mu}(p,p^{\prime})\,u(p) \:.
\end{equation}
By the Lorentz invariance and the current
conservation we get for $Q^2=-q^2$
\begin{equation}
{\cal
O}^{\mu}(p,p^{\prime})=\gamma^{\mu}F_1(Q^2)
+i\frac{\sigma^{\mu\nu}}{2M_N}q_{\nu}
F_2(Q^2)\,,
\end{equation}
where $M_N\simeq940~{\rm MeV}$ is the nucleon mass. 
From the effective action (\ref{action}) we can easily obtain the EM form factors, given as the overlap of the wave functions in the holographic direction~\cite{Hong:2007dq}: the Dirac and Pauli form factors are given respectively as
\begin{eqnarray}
F_{1}(Q^2)&=&
\sum_{k=1}^{\infty} \left(g^{(k)}_{V,min}Q_{\rm em}
+g_{V,mag}^{(k)}\,I_3\right)\frac{\zeta_k m_{2k+1}^2}
{Q^2+m_{2k+1}^2}\,,
\label{formf}\\
F_2(Q^2) &=&F_2^3(Q^2)\,I_3= \,I_3\sum_{k=1}^{\infty}
{g_2^{(k)}\zeta_n m_{2k+1}^2\over Q^2+m_{2k+1}^2}\,,
\end{eqnarray} 
where $m_{2k+1}$ is the mass of $k$-th (proper) vector meson and  $g_V^{(k)}$, $\zeta_k$ are given by the wave function overlap of the nucleons and  $k$-th (proper) vector meson. 
The electromagnetic mass of nucleons become in the Landau gauge, which is consistent with the Ward-Takahashi identity for the $\gamma$ matrix structure of the vertex in our approximation~\cite{Miransky:1994vk}, 
\begin{equation}
\delta M_{p\,(n)}=e^2\int\frac{{\rm d}^4Q}{(2\pi)^4}\left[F_1^{p\,(n)}(Q^2)\right]^2\frac{3M_N}{Q^2+M_N^2}\cdot\frac{1}{Q^2}\,.
\end{equation}
We find, keeping the first four vector mesons ($k=1,\cdots,4$) in the form factors~\cite{Hong:2007dq}, that 
\begin{equation}
\delta M_p=0.494~{\rm MeV},\quad \delta M_n=0.018~{\rm MeV}. 
\end{equation}
The holographic estimate of EM mass difference of nucleons, $\Delta_{\rm QED}M_N\equiv\delta M_p-\delta M_n=0.48\,{\rm MeV}$ compares roughly well with recent lattice calculations~~\cite{Blum:2007cy,Basak:2008na,Blum:2010ym,deDivitiis:2013xla,Borsanyi:2013lga,Borsanyi:2014jba}. %, $\Delta_{\rm QED}M_N=1.59\pm0.30\pm0.35\,({\rm MeV})$\,~\cite{Borsanyi:2013lga}.

To conclude we have calculated the electromagnetic contributions to the hadron mass in the Sakai-Sugimoto model of holographic QCD, where the Kaluza-Klein scale is taken to be $M_{\rm KK}=949~{\rm MeV}$, setting the scale of our estimate of EM corrections. For pions, the vacuum alignment due to the electromagnetic interactions has been calculated holographically to obtain the mass difference $m_{\pi^{\pm}}-m_{\pi^0}=1.8~{\rm MeV}$, which is less than half of the experimental value, $4.5~{\rm MeV}$.   The EM mass of pions is finite because of the Weinberg sum rules on spectral functions. Numerically we have shown that the EM mass is finite. We then calculate the electromagnetic mass of nucleons in the effective theory of holographic baryons, derived from the Sakai-Sugimoto model and find the EM mass of proton $\delta M_p=0.494~{\rm MeV}$ and that of neutron $\delta M_n=0.018~{\rm MeV}$ to get $\left(M_p-M_n\right)_{\rm QED}=0.48~{\rm MeV}$\,. Our estimate of EM mass difference of nucleons agrees with recent lattice calculations, though the latest lattice result~\cite{Borsanyi:2014jba} prefers  a bit larger value than ours. 

%%%%%%%%%%%%%%%%%%%%%%%%%%%%%%%%%%%%%%%%%%%%%%%%%%%%%%%%%%%%%%%%%%%%
%\eject
\subsection{Acknowledgements}
%\acknowledgments 
%This work (D.~K.~H.) was supported for two years by Pusan National University Research grant. 
%The work of D.~K.~H. was supported by the Korea Research Foundation Grant funded by the Korean Government (KRF-2008-341-C00008).
We wish to thank  J.-W. Chen, J. Harvey, K. Hashimoto, J. Sonnenschein, S. Sugimoto, H.-U. Yee for useful discussions and C. Kim for the help with mathematica. The author is especially grateful to  S. Krieg and K.\,K. Szabo for informing him of  their very recent lattice results~\cite{Borsanyi:2014jba}.  This research was supported by Basic Science Research Program through the National Research Foundation of Korea (NRF) funded by the Ministry of Education (NRF-2013R1A1A2011933). We acknowledge the hospitality at APCTP and also at IPMU where part of this work was done.


\begin{thebibliography}{99}

%\cite{Sakai:2004cn}
\bibitem{Sakai:2004cn} 
  T.~Sakai and S.~Sugimoto,
  %``Low energy hadron physics in holographic QCD,''
  Prog.\ Theor.\ Phys.\  {\bf 113}, 843 (2005)
  [hep-th/0412141].
  %%CITATION = HEP-TH/0412141;%%
  %824 citations counted in INSPIRE as of 23 Jul 2014
 
 
  %\cite{Sakai:2005yt}
\bibitem{Sakai:2005yt} 
  T.~Sakai and S.~Sugimoto,
  %``More on a holographic dual of QCD,''
  Prog.\ Theor.\ Phys.\  {\bf 114}, 1083 (2005)
  [hep-th/0507073].
  %%CITATION = HEP-TH/0507073;%%
  %488 citations counted in INSPIRE as of 27 Aug 2014 
  
  %\cite{Erlich:2005qh,Da Rold:2005zs,Hong:2006ta}
\bibitem{Erlich:2005qh} 
  J.~Erlich, E.~Katz, D.~T.~Son and M.~A.~Stephanov,
  %``QCD and a holographic model of hadrons,''
  Phys.\ Rev.\ Lett.\  {\bf 95}, 261602 (2005)
  [hep-ph/0501128].
  %%CITATION = HEP-PH/0501128;%%
  %631 citations counted in INSPIRE as of 23 Jul 2014
  
  %\cite{Da Rold:2005zs}
\bibitem{Da Rold:2005zs} 
  L.~Da Rold and A.~Pomarol,
  %``Chiral symmetry breaking from five dimensional spaces,''
  Nucl.\ Phys.\ B {\bf 721}, 79 (2005)
  [hep-ph/0501218].
  %%CITATION = HEP-PH/0501218;%%
  %491 citations counted in INSPIRE as of 27 Aug 2014
  
  
  %\cite{Hong:2006ta}
\bibitem{Hong:2006ta} 
  D.~K.~Hong, T.~Inami and H.~U.~Yee,
  %``Baryons in AdS/QCD,''
  Phys.\ Lett.\ B {\bf 646}, 165 (2007)
  [hep-ph/0609270].
  %%CITATION = HEP-PH/0609270;%%
  %97 citations counted in INSPIRE as of 27 Aug 2014
 

 
 %lattice
 %\cite{Blum:2007cy,Basak:2008na,Blum:2010ym,deDivitiis:2013xla,Borsanyi:2013lga}
\bibitem{Blum:2007cy} 
  T.~Blum, T.~Doi, M.~Hayakawa, T.~Izubuchi and N.~Yamada,
  %``Determination of light quark masses from the electromagnetic splitting of pseudoscalar meson masses computed with two flavors of domain wall fermions,''
  Phys.\ Rev.\ D {\bf 76}, 114508 (2007)
  [arXiv:0708.0484 [hep-lat]].
  %%CITATION = ARXIV:0708.0484;%%
  %48 citations counted in INSPIRE as of 27 Aug 2014
 
 %\cite{Basak:2008na}
\bibitem{Basak:2008na} 
  S.~Basak {\it et al.}  [MILC Collaboration],
  %``Electromagnetic splittings of hadrons from improved staggered quarks in full QCD,''
  PoS LATTICE {\bf 2008}, 127 (2008)
  [arXiv:0812.4486 [hep-lat]].
  %%CITATION = ARXIV:0812.4486;%%
  %28 citations counted in INSPIRE as of 27 Aug 2014
 
 %\cite{Blum:2010ym}
\bibitem{Blum:2010ym} 
  T.~Blum, R.~Zhou, T.~Doi, M.~Hayakawa, T.~Izubuchi, S.~Uno and N.~Yamada,
  %``Electromagnetic mass splittings of the low lying hadrons and quark masses from 2+1 flavor lattice QCD+QED,''
  Phys.\ Rev.\ D {\bf 82}, 094508 (2010)
  [arXiv:1006.1311 [hep-lat]].
  %%CITATION = ARXIV:1006.1311;%%
  %77 citations counted in INSPIRE as of 27 Aug 2014
  
  %\cite{deDivitiis:2013xla}
\bibitem{deDivitiis:2013xla} 
  G.~M.~de Divitiis {\it et al.}  [RM123 Collaboration],
  %``Leading isospin breaking effects on the lattice,''
  Phys.\ Rev.\ D {\bf 87}, no. 11, 114505 (2013)
  [arXiv:1303.4896 [hep-lat]].
  %%CITATION = ARXIV:1303.4896;%%
  %26 citations counted in INSPIRE as of 27 Aug 2014
 
 
 %\cite{Borsanyi:2013lga}
\bibitem{Borsanyi:2013lga} 
  S.~Borsanyi, S.~Dürr, Z.~Fodor, J.~Frison, C.~Hoelbling, S.~D.~Katz, S.~Krieg and T.~Kurth {\it et al.},
  %``Isospin splittings in the light baryon octet from lattice QCD and QED,''
  Phys.\ Rev.\ Lett.\  {\bf 111}, 252001 (2013)
  [arXiv:1306.2287 [hep-lat]].
  %%CITATION = ARXIV:1306.2287;%%
  %19 citations counted in INSPIRE as of 27 Aug 2014
 
 %\cite{Borsanyi:2014jba}
\bibitem{Borsanyi:2014jba} 
  S.~Borsanyi, S.~Durr, Z.~Fodor, C.~Hoelbling, S.~D.~Katz, S.~Krieg, L.~Lellouch and T.~Lippert {\it et al.},
  %``Ab initio calculation of the neutron-proton mass difference,''
  arXiv:1406.4088 [hep-lat].
  %%CITATION = ARXIV:1406.4088;%%
  %9 citations counted in INSPIRE as of 09 Oct 2014
 
 
 
  
 %\cite{Dashen:1970et}
\bibitem{Dashen:1970et} 
  R.~F.~Dashen,
  %``Some features of chiral symmetry breaking,''
  Phys.\ Rev.\ D {\bf 3}, 1879 (1971).
  %%CITATION = PHRVA,D3,1879;%%
  %270 citations counted in INSPIRE as of 24 Jul 2014 
  
%\cite{Weinberg:1975gm}
\bibitem{Weinberg:1975gm} 
  S.~Weinberg,
  %``Implications of Dynamical Symmetry Breaking,''
  Phys.\ Rev.\ D {\bf 13}, 974 (1976).
  %%CITATION = PHRVA,D13,974;%%
  %1271 citations counted in INSPIRE as of 24 Jul 2014
  
  
 %\cite{Peskin:1980gc,Preskill:1980mz}
\bibitem{Peskin:1980gc} 
  M.~E.~Peskin,
  %``The Alignment of the Vacuum in Theories of Technicolor,''
  Nucl.\ Phys.\ B {\bf 175}, 197 (1980).
  %%CITATION = NUPHA,B175,197;%%
  %375 citations counted in INSPIRE as of 02 Sep 2014
  
 %\cite{Preskill:1980mz}
\bibitem{Preskill:1980mz} 
  J.~Preskill,
  %``Subgroup Alignment in Hypercolor Theories,''
  Nucl.\ Phys.\ B {\bf 177}, 21 (1981).
  %%CITATION = NUPHA,B177,21;%%
  %262 citations counted in INSPIRE as of 02 Sep 2014  
  
  %\cite{Aharony:2008an,Hashimoto:2008sr,Hong:2007tf}
\bibitem{Aharony:2008an} 
  O.~Aharony and D.~Kutasov,
  %``Holographic Duals of Long Open Strings,''
  Phys.\ Rev.\ D {\bf 78}, 026005 (2008)
  [arXiv:0803.3547 [hep-th]].
  %%CITATION = ARXIV:0803.3547;%%
  %50 citations counted in INSPIRE as of 30 Aug 2014
  
  %\cite{Hashimoto:2008sr}
\bibitem{Hashimoto:2008sr} 
  K.~Hashimoto, T.~Hirayama, F.~L.~Lin and H.~U.~Yee,
  %``Quark Mass Deformation of Holographic Massless QCD,''
  JHEP {\bf 0807}, 089 (2008)
  [arXiv:0803.4192 [hep-th]].
  %%CITATION = ARXIV:0803.4192;%%
  %28 citations counted in INSPIRE as of 30 Aug 2014
  
  %\cite{Hong:2007tf}
\bibitem{Hong:2007tf} 
  D.~K.~Hong, H.~-C.~Kim, S.~Siwach and H.~-U.~Yee,
  %``The Electric Dipole Moment of the Nucleons in Holographic QCD,''
  JHEP {\bf 0711}, 036 (2007)
  [arXiv:0709.0314 [hep-ph]].
  %%CITATION = ARXIV:0709.0314;%%
  %16 citations counted in INSPIRE as of 24 Jul 2014
 
 %\cite{Hong:2004sa}
\bibitem{Hong:2004sa} 
  S.~Hong, S.~Yoon and M.~J.~Strassler,
  %``On the couplings of vector mesons in AdS / QCD,''
  JHEP {\bf 0604}, 003 (2006)
  [hep-th/0409118].
  %%CITATION = HEP-TH/0409118;%%
  %81 citations counted in INSPIRE as of 10 Oct 2014
 
 %\cite{Weinberg:1996kr}
\bibitem{Weinberg:1996kr} 
  S.~Weinberg,
  ``The quantum theory of fields. Vol. 2: Modern applications,''
  Cambridge, UK: Univ. Pr. (1996) 489 p.
  %59 citations counted in INSPIRE as of 17 Dec 2014
 
 %\cite{Das:1967it}
\bibitem{Das:1967it} 
  T.~Das, G.~S.~Guralnik, V.~S.~Mathur, F.~E.~Low and J.~E.~Young,
  %``Electromagnetic mass difference of pions,''
  Phys.\ Rev.\ Lett.\  {\bf 18}, 759 (1967).
  %%CITATION = PRLTA,18,759;%%
  %325 citations counted in INSPIRE as of 02 Sep 2014
  
  
 %\cite{Witten:1979kh}
\bibitem{Witten:1979kh} 
  E.~Witten,
  %``Baryons in the 1/n Expansion,''
  Nucl.\ Phys.\ B {\bf 160}, 57 (1979).
  %%CITATION = NUPHA,B160,57;%%
  %2027 citations counted in INSPIRE as of 02 Sep 2014


%\cite{Hong:2008nh}
\bibitem{Hong:2008nh} 
  D.~K.~Hong, K.~M.~Lee, C.~Park and H.~U.~Yee,
  %``Holographic Monopole Catalysis of Baryon Decay,''
  JHEP {\bf 0808}, 018 (2008)
  [arXiv:0804.1326 [hep-th]].
  %%CITATION = ARXIV:0804.1326;%%
  %10 citations counted in INSPIRE as of 12 Sep 2014 
  
  
%\cite{Hong:2007kx,Hong:2007ay,Hata:2007mb} 
 %\cite{Hong:2007kx}
\bibitem{Hong:2007kx} 
  D.~K.~Hong, M.~Rho, H.~U.~Yee and P.~Yi,
  %``Chiral Dynamics of Baryons from String Theory,''
  Phys.\ Rev.\ D {\bf 76}, 061901 (2007)
  [hep-th/0701276 [HEP-TH]]:
  %%CITATION = HEP-TH/0701276;%%
  %146 citations counted in INSPIRE as of 03 Sep 2014
  
  
 %\cite{Hong:2007ay}
\bibitem{Hong:2007ay} 
D.~K.~Hong, M.~Rho, H.~U.~Yee and P.~Yi,
  %``Dynamics of baryons from string theory and vector dominance,''
  JHEP {\bf 0709}, 063 (2007)
  [arXiv:0705.2632 [hep-th]].
  %%CITATION = ARXIV:0705.2632;%%
  %110 citations counted in INSPIRE as of 03 Sep 2014  
   
   
%\cite{Hata:2007mb}
\bibitem{Hata:2007mb} 
  H.~Hata, T.~Sakai, S.~Sugimoto and S.~Yamato,
  %``Baryons from instantons in holographic QCD,''
  Prog.\ Theor.\ Phys.\  {\bf 117}, 1157 (2007)
  [hep-th/0701280 [HEP-TH]].
  %  %CITATION = HEP-TH/0701280;%%
  %179 citations counted in INSPIRE as of 16 Sep 2014   
   
 %\cite{Hashimoto:2010je}
\bibitem{Hashimoto:2010je} 
  K.~Hashimoto, N.~Iizuka and P.~Yi,
  %``A Matrix Model for Baryons and Nuclear Forces,''
  JHEP {\bf 1010}, 003 (2010)
  [arXiv:1003.4988 [hep-th]].
  %%CITATION = ARXIV:1003.4988;%%
  %7 citations counted in INSPIRE as of 03 Sep 2014
 
 


 
 
 %\cite{Hong:2007dq}
\bibitem{Hong:2007dq} 
  D.~K.~Hong, M.~Rho, H.~U.~Yee and P.~Yi,
  %``Nucleon form-factors and hidden symmetry in holographic QCD,''
  Phys.\ Rev.\ D {\bf 77}, 014030 (2008)
  [arXiv:0710.4615 [hep-ph]].
  %%CITATION = ARXIV:0710.4615;%%
  %74 citations counted in INSPIRE as of 23 Sep 2014

%\cite{Miransky:1994vk}
\bibitem{Miransky:1994vk} 
  V.~A.~Miransky,
  ``Dynamical symmetry breaking in quantum field theories,''
  Singapore, Singapore: World Scientific (1993) 533 p.
  %2 citations counted in INSPIRE as of 29 Sep 2014

 
 \end{thebibliography}
\end{document}